# Using Mobilize Power Management IP for Dynamic & Static Power Reduction in SoC at 130 nm

Dan Hillman
*Virtual Silicon*
danh@virtual-silicon.com

**Abstract**

*At 130 nm and 90 nm, power consumption (both dynamic and static) has become a barrier in the roadmap for SoC designs targeting battery powered, mobile applications. This paper presents the results of dynamic and static power reduction achieved implementing Tensilica's 32-bit Xtensa microprocessor core, using Virtual Silicon's Power Management IP. Independent voltage islands are created using Virtual Silicon's VIP PowerSaver standard cells by using voltage level shifting cells and voltage isolation cells to implement power islands. The VIP PowerSaver standard cells are characterized at 1.2V, 1.0V and 0.8V, to accommodate voltage scaling. Power islands can also be turned off completely. Designers can significantly lower both the dynamic power and the quiescent or leakage power of their SoC designs, with very little impact on speed or area using Virtual Silicon's VIP Gate Bias standard cells.*

## 1.0 Overview

Designers today are challenged not only by new process technologies capable of incorporating more and more devices on a single chip, but also managing the increase in the power that goes along with it. Techniques such as clock gating, low power processes, low power IP and lower supply voltage used with each new generation of process technology have helped designers of mobile applications to stem the tide of ever increasing power. Configurable processors such as the Xtensa [1] core used here can also optimize the instruction set to minimize power and energy.

As technology moved down to 180nm, leakage power was not a major issue. The old techniques produced acceptable results. However, the130nm and 90nm technologies are reaching the point where leakage power is nearly the same as dynamic power. And both dynamic and static powers are increasing at such a rate that the old rules just don't work. Unless new techniques are found, 130nm and 90nm technologies for handheld devices will be adopted more slowly.

Significantly reduce both dynamic and static power of SoC using Virtual Silicon Mobilize Power Management IP. A 32-bit Xtensa microprocessor will be synthesized and taken all the way through routing in order to benchmark the power reduction in dynamic power and static (leakage) power.

## 2.0 Dynamic power reduction

Dynamic or switching power is expressed by $P_{dynamic} = k*C*V^2*F*SA$: where), $k$ = Constant (usually varies from 0 to 1), C represents capacitance, V is the operational voltage and F is the frequency for the design, and SA = Switching activity.

The configurable Xtensa attacks most variables in the dynamic power equation to achieve low power designs, e.g., smaller microprocessor configuration without unused features hence reduces C, advanced Xtensa architectures allow more tasks get done within each clock cycle or at minimum F, and Xtensa's extensive clock gating makes possible the lowest SA. For a detailed presentation see [2]. This work focuses on minimizing the operating voltage to reduce dynamic power. The libraries for most of today's designs are characterized for small process variations at a single voltage. The performance is guaranteed for process, temperature and voltage variations. What if you could get a library that was characterized for several voltages? You could then design each part of the SoC at its minimum voltage for its required operating frequency.

### 2.1 Design overview

The example design shown in Figure 1 is made up of three components: an Xtensa processor, a memory block and a USB core.

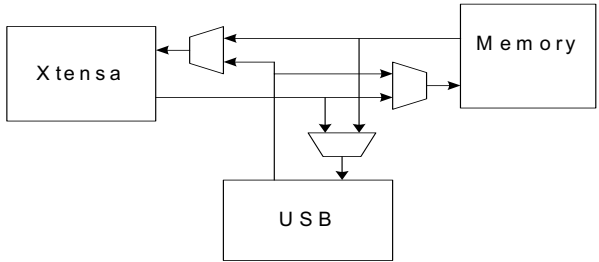

**Figure 1: Simple SoC**

We divided the design into three autonomous power islands, Xtensa, memory and USB. The goal was to get the minimum system dynamic power by running each power island at its minimum required voltage.

For a 512K embedded SRAM, commercial memory compilers can deliver the access times such that the critical paths between Xtensa and memory can run up to 181MHz at 0.8V worst case TSMC G process. The USB is a hard core and it requires a supply of 1.2V. The minimum processor system performance is 150MHz at the 130nm node [3], which the designer wants to obtain at the lowest voltage possible. To find the minimum operating voltage for the Xtensa core at 150MHz, we used Virtual Silicon's PowerSaver library. The library is characterized at three voltages, 1.2V, 1.0V and 0.8V.

### 2.2 Design flow

Figure 2 shows the design flow used to find the Xtensa's minimum operating voltage at 150MHz.

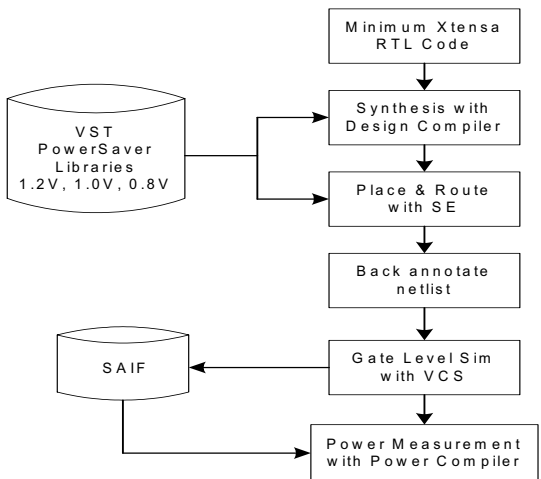

**Figure 2: Design Flow**

The Xtensa was synthesized at 1.2V, 1.0V and 0.8V using worst case wireload corner .db in DC. The best case corner was used for hold time fixing in SE. Routing used 5 layers in a 6 layer system. Post route simulation was used to create SAIF files for measuring power in Power Compiler.

### 3.0 Synthesis and layout results

For each voltage the design was synthesized and routed at 167MHz (6.0ns), 182MHz (5.5ns), 200MHz (5.0ns), 222MHz (4.5ns), 250MHz (4.0ns), 286MHZ (3.5ns), 333MHz (3.0ns) and 400MHz (2.5ns). The goal for the design is 150MHz post layout. The results for the 1.2V, 1.0V and 0.8V libraries after place and route are shown in Figure 3.

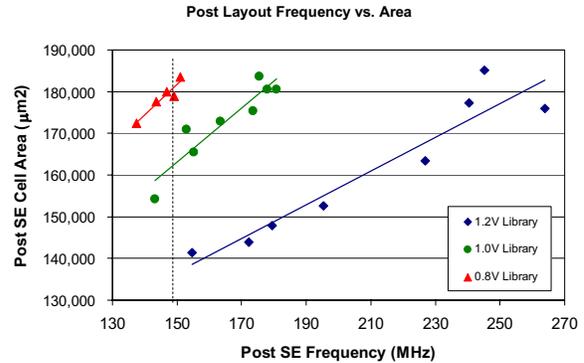

**Figure 3: Post Route Results**

Not surprisingly, for the 1.2V library all of the synthesis and post SE runs meet timing. The utilization for this set of designs was 80%. Higher utilization resulted in too many DRC violations. The area curve generally looks as you would expect. A faster design yields a larger area. The best area result that still met post route timing was for the 167MHz design. Actual post layout clock was 155MHz, and the post route area was 141,429 µm2.

For the 1.0V library, all but the 167MHz constrained design met the required timing. The utilization for this set of designs was 80% to begin with; the post layout ending utilization is 89%. Higher utilization resulted in too many DRC violations. The best result that still met timing was for the 182MHz design. Actual post layout clock was 155MHz and the post route area was 165,424 µm2.

For the 0.8V library, only the 333MHz constrained design met the required timing. Synthesis and routing for 167MHz and 182MHz designs were not done. The initial utilization for this set of designs was 75%; the post layout final utilization was 83%. Higher utilization resulted in too many DRC violations. Actual post layout clock was 151Mhz and the post route area was 183,551 µm2.

A summary of the post route results vs. the library voltages for the smallest area design that still met timing is shown in Figure 4.

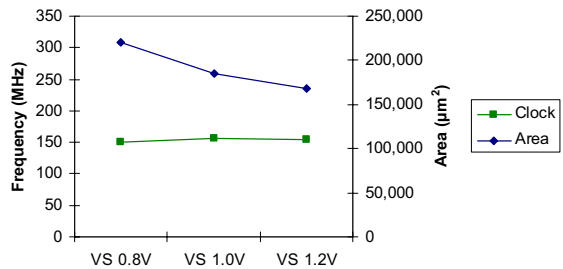

**Figure 4: Library Voltage vs. Area**

The minimum required performance of 150MHz was met by all three libraries. For the 1.0V and 0.8V libraries performance was achieved at the cost of increasing the design area. Since the design is small the increase in area is worth the power reduction obtained.

## 4.0 Dynamic power simulation results

Our goal was to minimize the dynamic power used by the Xtensa core. To measure dynamic power, we took the post layout back annotated netlist and ran six different Verilog test suites to generate SAIF files. The SAIF files were fed into Power Compiler to generate dynamic power numbers. The results are shown in Figure 5.

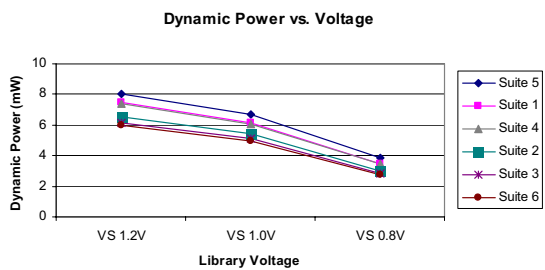

**Figure 5: Dynamic Power vs. Library Voltage**

As expected, you can see dynamic power varied by suite. Compared to the 1.2V design, the 1.0V design reduced power by 17% while increasing area by 10%. The maximum theoretical power reduction for a design going from 1.2V to 1.0V is 30.5% $((1.2^2 – 1.0^2)/1.2^2)$. The actual dynamic power savings was less than the theoretical results because of the increased capacitance from the extra gates and routing area.

Compared to the 1.2V design, the 0.8V design reduces power by 53% while increasing area by 31%. The maximum theoretical power reduction for a design going from 1.2V to 0.8V is 55.5% $((1.2^2 – 0.8^2)/1.2^2)$. Again, the actual dynamic power savings was less than the theoretical results because of the increased capacitance from the extra gates and routing area.

## 5.0 Putting the soc together

The islands of the SoC have been optimized for their minimum voltage for proper operation. Xtensa and the memory can operate at 0.8V and the USB can operate at 1.2V. Level shifters will be needed to ensure proper operation for the SoC as a whole. One way to employ level shifters is be to put a level shifter on all output signals that are less than 1.2V. In the SoC shown in Figure 6, all the muxes then operate off a common 1.2V supply.

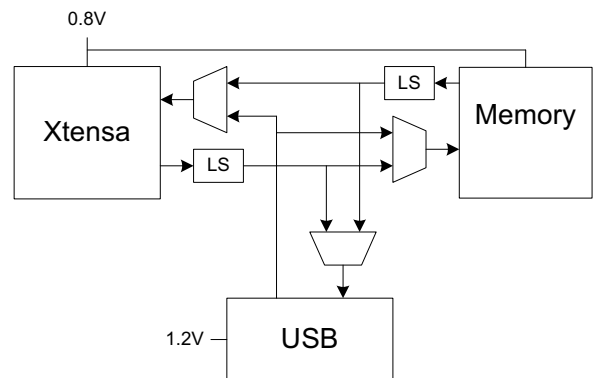

**Figure 6: Simple SoC with Level Shifters**

Down shifters are not required because in the Virtual Silicon library, all the inputs to cells drive gates. Therefore gates tied to a VDD less than 1.2V can safely be driven to 1.2V without damage. However, if your library has transmission gates, down shifters may be required to prevent unwanted currents or latch up.

Some optimization of level shifter usage can also be done. Level shifters are not needed for signals that only go between the memory the Xtensa. Likewise, level shifters are not needed for signals that only go from the USB to the Xtensa or memory.

## 6.0 Dynamic power reduction summary

We were able to achieve our performance goal of 150MHz post-route performance at 0.8V. Our simple SoC can be divided into two power islands; USB at 1.2V, memory and Xtensa at 0.8V. This voltage reduction on the Xtensa power island netted a 53% power reduction with the Mobilize IP. The level shifters and voltage isolation gates allowed us to easily create power islands in order to realize this dynamic power reduction in our SoC. Multi voltage designs using power islands created with Molbilize IP can be implemented.

## 7.0 Static (leakage) power reduction

Power reduction techniques such as voltage scaling and clock gating will work well for dynamic power, but not for leakage power. Leakage power for 130nm and 90nm is approaching that of dynamic power. This section will deal with incorporating static (leakage power) reduction in addition to dynamic power reduction

## 7.1 Device leakage mechanisms

To minimize leakage power it important to understand the leakage sources in a device. Figure 7 is a diagram of those leakages.

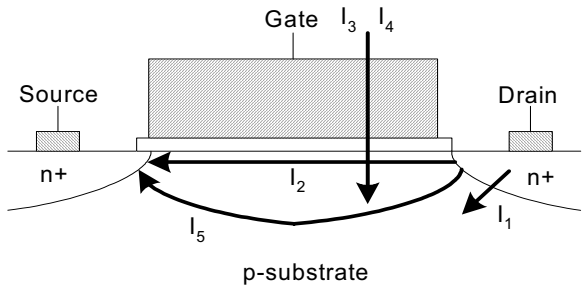

**Figure 7: Leakage Mechanisms**

I1 is the reverse bias junction leakage from the drain to the substrate. This leakage is about two to three orders of magnitude less than the other leakage mechanisms. I2 is sub-threshold leakage. This is the leakage when the gate to source voltage is zero volts. For 130nm and 90nm this is the major source of leakage power. I3 is leakage through the gate oxide to the substrate, source and drain of the device. For 130nm and 90nm the gate oxide is only several atoms thick. At 90nm this leakage is two orders of magnitude less than gate leakage. I4 hot-carrier injection and I5 off state leakage are both minor sources of leakage.
Shown in Table 1 is how the different device leakage mechanisms change with process.

|  | 180nm | 130nm | 90nm |
|---|---|---|---|
| I1 = reverse bias junction | minor | minor | minor |
| I2 = sub-threshold | minor | major | major+ |
| I3 = gate oxide tunneling | minor | relevant | significant |
| I4 = hot-carrier injection | minor | minor | minor |
| I5 = off state leakage | minor | minor | minor |

**Table 1: Leakage vs. Process**

## 8.0 Gate Bias, what is it?

Virtual Silicon has developed a standard cell library that puts an NMOS sleep transistor into the path to ground with SLPB as the sleep gate bias. The concept of sleep devices, both PMOS and NMOS, has been around for some time. What makes this library unique is that a sleep device is in every library cell and the sleep gate is driven to a negative voltage. Figure 8 shows how these devices are connected in two different cells.

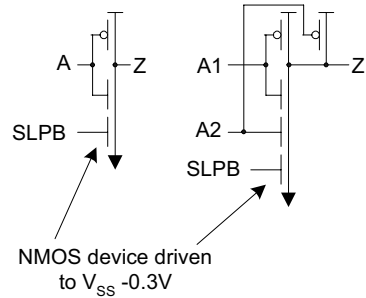

**Figure 8: Inverter and NAND Gate with Sleep Device**

The sleep device is a regular threshold device and not a high threshold device. Because this library is built on a standard process more IP is available and the wafer costs are less than for a low power process.

Placing the sleep device in every cell has a number of benefits. It is easy to sleep small sections of logic for example, an ALU. The sleep signal may be turned on be an already existing global clock gating signal. By using a clock gating signal, not only does clock stop to reduce dynamic power, but also the logic is put sleep to reduce static power. Also, as compared to one large sleep device the sizing and placement of the sleep device is done for you. The Gate Bias technology allows the creation of data retention flip-flops for fast save and restore operation. There is only a minor impact on speed because the sleep device is NMOS.
Figure 9 shows the leakage reduction obtained by driving the NMOS gate voltage negative.

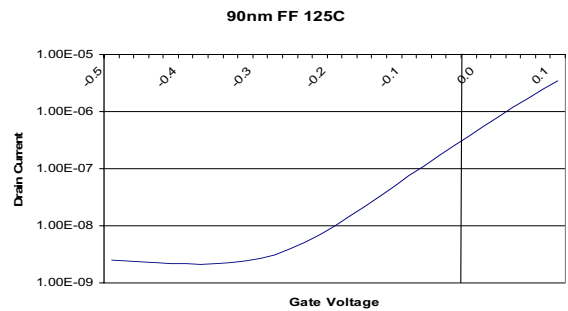

**Figure 9: NMOS Leakage Current vs. Gate Voltage**

At zero volts on the gate, the leakage current is about 0.5μA. At a sleep voltage of -0.3V, the leakage current is reduced by 260x (to about 2nA). The sub-threshold leakage, which is the dominant leakage, is effectively lowered by this technique.

## 8.1 Example design

Figure 10 shows an example of small SoC with CPU, memory and a logic block. In Figure 10, LS stand for level shifter and ISO stands for isolation gate. For purposes of this discussion we will optimize the logic block for minimum leakage using the Gate Bias library. The minimum frequency requirement for the logic block is 200MHz.

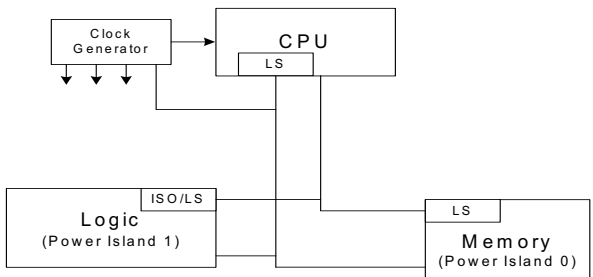

**Figure 10: Example SoC**

## 9.0 Design Flow

RTL coding should keep all modules of an island in one hierarchy. Level shifters and isolation gates should also be instantiated at the top level of the module.

Synthesis (DC) does not understand sleep functionality. In order for synthesis to use this library the sleep functionality must first be removed from the library. The Virtual Silicon library provides a .libsyn and .dbsyn files without the sleep functionality. As shown in Figure 11, .dbsyn is used by DC to create a netlist without sleep. A script is used to add the sleep pin to the netlist after synthesis

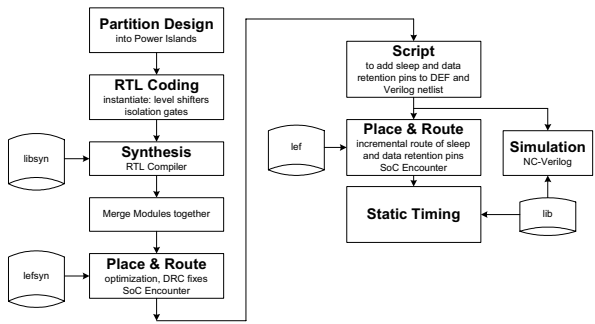

**Figure 11: Gate Bias Design Flow**

Since there is no RTL construct for sleep, the sleep functionality can only be simulated at the gate level. A script was used to add the sleep pin to the netlist. The simulator uses gate level primitives which has the sleep functionality in it.

## 9.1 Synthesis

The 200MHz minimum frequency requirement for the logic module could only be met at 1.2V supply.

## 10.0 Managing static power

Now that we have a design with the sleep device in it, we need to generate the negative voltage for sleep and control the isolation gates. The Virtual Silicon Power Island Manager is IP that generates the negative voltage for sleep, controls going in and out of sleep, isolation gates, and data retention flip-flops. The Power Island Manager can be controlled by the CPU with write and read commands. Figure 12 shows the Power Island Manager connected in the design.

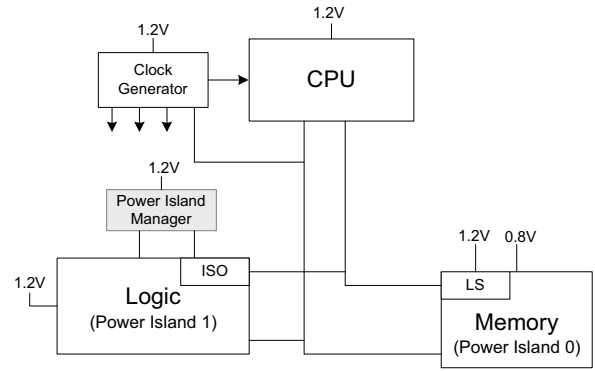

**Figure 12: Example SoC with Power Management**

To put the logic to sleep the CPU writes to the sleep bit in the Power Island Manager. The Power Island Manager has a state machine that controls the isolation signal, sleep signal, and data retention signal. The state machine ensures the timing of these signals to give proper operation to the design. To come out of the sleep state the CPU writes the sleep bit. A status bit in the Power Island Manager can be read by the CPU to know when the logic is out of the sleep state and ready to resume operation. Times enter or leave sleep are approximately 60ns.

Although the CPU could also be put to sleep, presentation on this topic is beyond the scope of this paper.

## 10.1 Static power reduction

The logic block contains 108,000 gates and needs a Power Island Manager to bring it in and out of sleep. For the 'G' process, 108,000 typically consume 208μW. With the Gate Bias library the typical power for the 108,000 gates and the Power Island Manager is 4.9μW. Figure 13 shows the power comparison for 'G' and Gate Bias library. While in sleep, the Gate Bias technology saves 97.6% of the static power.

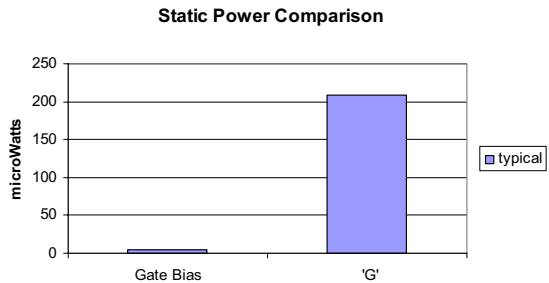

**Figure 13 Static Power Comparison**

### 10. 2 Static power reduction summary

We have shown a 97.6% static power reduction using the Gate Bias library and Power Management IP.

## 11.0 Test chip results

A Gate Bias test chip containing the standard cells, memory and I/O has been fabricated and evaluated.

The standard cells got even better leakage reduction than predicted by the Spice models. Results for the two input Nan gate are shown in Figures 14 and 15.

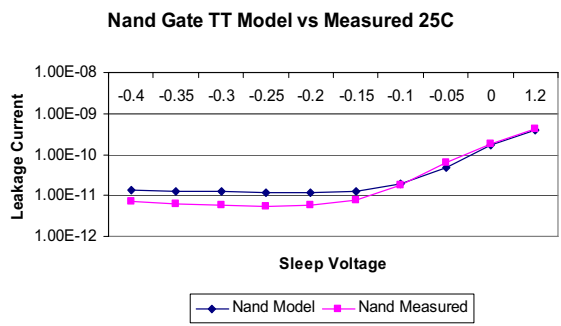

**Figure 14 Model vs. Measured 25ºC**

At 25ºC, the model predicts a leakage reduction of 33.9x and the measure result gives a leakage reduction of 78.6x.

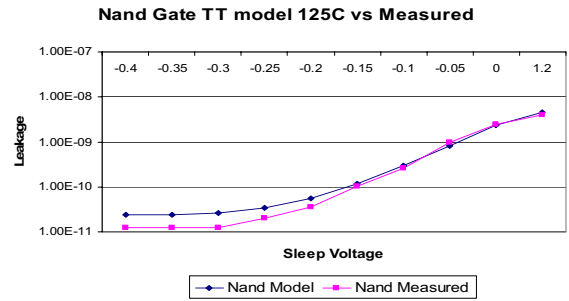

**Figure 15 Model vs. Measured 125ºC**

At 125ºC, the model predicts a leakage reduction of 197x and the measure result gives a leakage reduction of 326x.

The on-chip negative bias generator was proven to function as designed.

Gate Bias technology was added to SRAM to reduce leakage and retain data. A sleep and data retention signals were applied to the test chip SRAM. The data was retained down to 0.3V $V_{DD}$ supply. The leakage reduction at 125ºC is shown in Figure 16.

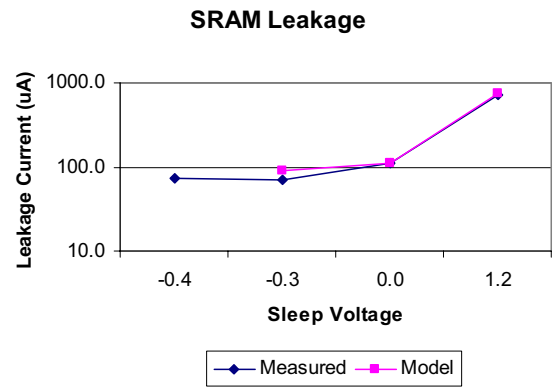

**Figure 16 SRAM Leakage Reduction**

At 125ºC, the model predicts a leakage reduction of 8.1x and the measure result gives a leakage reduction of 10.0x. The leakage current was reduced from 719μA to 71.5μA. Based on test chip results, an SRAM design has been done which will reduce the leakage from 20x to 65x depending on the SRAM size and configuration.

## 12.0 Summary

We have demonstrated how dynamic power has been cut by 53% by running the Xtensa processor at 0.8V, while still meeting performance goals using the Virtual Silicon Mobilize power management IP. Static power has been dramatically reduced by 97% using the Gate Bias library and Power Management IP.

The Gate Bias technology has been validated by test chip results to get better than the predicted leakage reduction.

## 13.0 References

1. Each configuration of the extensible, configurable Xtensa processor is created by the Xtensa Processor Generator from a unique designer-defined specification of both predefined configuration options and designer-defined instruction extensions. The generator can create a wide range of processor cores, including small task engines, high performance DSPs, media processors and network processors all sharing the common base Xtensa instruction set architecture. The configuration used in this paper has the minimum set of configuration options and includes only the base ISA elements without custom instruction extensions. The power minimization techniques discussed can be applied equally to any Xtensa processor configuration.

2. Peters, E., Taglieri, G., & Vemury, L. (2000). Low Power Synthesis Flow For a Configurable Core. Boston: SNUG.

3. International Technology Roadmap for Semiconductors, System Drivers. (2001). (p. 17).